\documentclass{article}
\usepackage{buckow}

\def\beq{\begin{equation}}
\def\eeq{\end{equation}}

\begin {document}
\begin{flushright}LPTENS-02/02, PAR-LPTHE 02-01
\end{flushright}
\def\titleline{Flat connections from flat gerbes%\newtitleline
}
\def\authors{
Arjan Keurentjes\1ad \2ad}
\def\addresses{
\1ad
Laboratoire de Physique Th\'eorique de l'\'Ecole Normale Sup\'erieure \\
24, rue Lhomond\\
75230 Paris Cedex 05, France\\
\2ad
Laboratoire de Physique Th\'eorique et des Hautes Energies\\
Universit\'e Pierre et Marie Curie (Paris VI)
4, place Jussieu\\
75252  Paris Cedex 05, France\\

{\tt e-mail arjan.keurentjes@lpt.ens.fr} \\
}
\def\abstracttext{We discuss some aspects of heterotic-Type I
duality. We focus on toroidal compactification, with special attention
for the topology of the gauge group, and the topology of the
bundle. We review the arguments leading to a
classification of $Spin(32)/Z_2$-bundles over tori, suitable for
string compactifications. A central role is played by $n$-gerbes with
connection, a generalization of bundles with connection.}
\large
\makefront

\section{Introduction}

One of the most remarkable string-dualities that has been proposed in
the past decade is the conjectured S-duality between the heterotic
$Spin(32)/Z_2$ string and the type I string. In spite of the agreement of the
massless spectrum and the low energy effective action, there are so
many differences in the formulation of the two theories that a duality
relation may seem rather implausible at first sight. Nevertheless, the
duality has passed so many detailed tests that it seems likely that
the 2 perturbative theories indeed are different limits of a single
underlying theory. 

This brief review mainly focusses on the work reported in
\cite{Keurentjes01b}. We will discuss aspects of the duality that will
be relevant for compactification of the 2 theories. The duality does
not depend on the compactification manifold, and we will restict to
toroidal compactification, although our techniques should be applicable to more complicated situations.

\section{Toroidal compactification}

Both theories have a Yang-Mills sector, and one should choose an
appropriate bundle over the $k$-torus $T^k$ one compactifies on.
To ensure that the string equations of motion are obeyed, we require
that this bundle is flat.

Next we should specify the structure group of the bundle. The type I
string theory has open strings with Chan-Paton factors taking 32
values, with manifest gauge symmetry (the adjoint
representation of) $O(32)$. The construction of the heterotic string
(in its bosonic formulation) involves a self-dual 16 dimensional
Euclidean lattice, fixing the topology of the gauge group to
$Spin(32)/Z_2$. This paradox was resolved in \cite{Sen98,
Witten98b}: The type I theory has an instanton breaking $O(32)$ to its
connected component $SO(32)$, while a solitonic particle transforms in
the spinor representation of $Spin(32)$. Hence these non-perturbative
states fix the structure group to $Spin(32)/Z_2$.

A flat $Spin(32)/Z_2$-bundle  on a $k$-torus is completely characterized by $k$
commuting holonomies $\Omega_i \in Spin(32)/Z_2$ ($i=1,\ldots,k$) for 
$k$ independent 1-cycles of the torus. There are many
distinct possibilities. Some examples:
\begin{itemize}
\item Pick all $\Omega_i$ in an Abelian subgroup of $Spin(32)$. This
is always allowed, and results in the ``standard'' and Narain
compactification schemes.
\item Let $\Omega_i$ be commuting elements of $Spin(32)/Z_2$ that do
not commute when lifted to $Spin(32)$. This is ``absence of vector
structure'': States in the vector representation $\mathbf{32}$ (which
is absent) would make such compactifications inconsistent
\cite{Bianchi, Bianchi2, Sen, Witten97}.
\item Choose $\Omega_1$ such that it has a centralizer with
non-simply connected semi-simple part $G$. Then choose two (or
more) remaining holonomies such that they commute in $G$, but their
lifts to the simply connected cover $\tilde{G}$ of $G$ do
not \cite{Borel, Kac, Keurentjes99, Witten97}.
\item Non-trivial triples, quadruples, quintuples of holonomies with
and without vector structure \cite{deBoer, Borel, Kac, Keurentjes99}.
\end{itemize}

There is yet another complication: In string theory not even all
$Spin(32)/Z_2$ bundles are allowed! This can be seen in the following
ways.

One of the results of \cite{Borel} is that flat principal
bundles over the 3-torus are characterized by 2 pieces of
information. First, there is the
topology of the bundle. Restricting to bundles
compatible with the group $Spin(32)/Z_2$, the topology can be
specified by a ``generalized Stiefel-Whitney class''
\cite{Witten97}. Second, there is the Chern-Simons
invariant
\beq
\frac{1}{16 \pi^2 h} \int_{T^3} (A\textrm{d}A + \frac{2}{3} A^3)
\eeq
As shown in \cite{Borel}, this invariant takes typically rational (and not
necessarily integer!) values. But in string theory one has the
anomaly-relation
\beq
H = \textrm{d}B + CS(A) - CS(\omega)
\eeq
relating the field strength for the NS 2--form to the Chern-Simons
forms of the gauge connection $A$ and the spin connection $\omega$. On
the $k$-torus, we set $\omega=0$, but a non-trivial Chern-Simons
form for the gauge connection seems incompatible with vanishing of $H$
\cite{deBoer}.

In the heterotic theory, there is a nice microscopic explanation for
this anomaly. In Narain compactification, the momenta take the form:
\begin{eqnarray}
k^I & = & q^I + w_i a_i^I \\
k_{iL,R} & = & \frac{n_i - q^I a^I_i - w_j a_i^I a_j^I/2}{R} \pm
\frac{w_i R}{2}
\end{eqnarray}
For winding states (non-zero $w_i$), the representation vector $k^I$
picks up contributions from the Wilson line $a_i^I$, resulting in
representations that would not be present in a (particle) gauge
theory. It is precisely these representations that pick up phases
related to the Chern-Simons invariant, and turn the compactification
into an inconsistent one \cite{deBoer}.

\section{T-duality to type II orientifolds}

Compactifying type I theory on a $k$-torus, and applying T-dualities
in all directions, one arrives at a type II theory on an orientifold
$T^k/Z_2$ where the $Z_2$ acts as reflection on all coordinates of the
$k$-torus (the action on the string states can be found in the
references). The resulting configurations of D-branes and orientifold
planes provide a geometrical description of the moduli space of flat
connections. This can be done for any compactification with
an $O(32)$ bundle \cite{Keurentjes00a}, but for bundles that cannot be
lifted to $Spin(32)/Z_2$, and those with fractional Chern-Simons
invariants (defined over sub-3-tori), we should expect inconsistencies. 

A cohomology analysis suggests the existence of the following
orientifold planes (see \cite{Bergman, Hanany00} and references therein).
\beq
\begin{tabular}{|c||c|c|c|c|}
\hline
O$p$ plane & group & $R_{p+1}$  & ${\cal B}$ & ${\cal C}$ \\
\hline
O$p^-$ & $O(2n)$ &  $-2^{p-4}$ & 0 & 0 \\
O$p^+$ & $Sp(n)$  & $+ 2^{p-4}$ & 1 & 0 \\
$\widetilde{\textrm{O}p}{}^-$ & $O(2n+1)$ & $1-2^{p-4}$ & 0 & 1\\
$\widetilde{\textrm{O}p}{}^+$ & $Sp(n)$ & $+ 2^{p-4}$ & 1 & 1 \\
\hline
\end{tabular}
\eeq
Here $R_{p+1}$ stands for D$p$-brane charge, and ${\cal B}$ and ${\cal
C}$ are discrete $Z_2$ fluxes, from the NS 2--form and the RR
$(5-p)$--form. The latter can only be defined when $p \leq 5$, and we
will henceforth assume that planes with non-trivial ${\cal C}$ charge
do not exist for $p > 5$ (see \cite{Bergman, Hyakutake} for a
different conclusion).

The first consistency requirement is then: Only use
orientifold planes that exist. This eliminates many of the
inconsistent bundles. For example, $\widetilde{\textrm{O}8}{}^-$ planes
would occur in the dual to compactification on a circle with a
holonomy in $O(32)$ with determinant $-1$, but neither the bundle, nor
$\widetilde{\textrm{O}8}{}^-$ planes are consistent in string theory.

A second consistency requirement is based on the observation that
the orientifold planes emit discrete NS-- and RR--fluxes. Consistency
requires us to patch these fluxes together in such a way that the
resulting configuration obeys the equations of motion. Again,
the easiest way to achieve this (while preserving supersymmetry), is
to require vanishing field strengths. In technical terms, the discrete
tensorfluxes should give rise to flat $n$-gerbes, away from
the orientifold planes.

\section{The holonomy of flat $n$-gerbes}

The concept of a bundle can be generalized to that of an
$n$-gerbe. As a matter of fact, $0$-gerbes are just unitary line
bundles \cite{Hitchin}. We take the point of view of \cite{Hitchin},
and define a gerbe on a manifold in terms of transition functions over
overlapping patches, satisfying suitable cocycle conditions. 

A connection on an $n$-gerbe is an $(n+1)$-form $C_{n+1}$. Its gauge
transformation is
\beq
C_{n+1} \sim C_{n+1} + \textrm{d}C_{n}
\end{equation}
This is important when gluing the patches together: connections on
different patches may differ by a gauge transformation. The $C_{n}$
themselves are only defined up to an exact form; 
\beq
C_{n} \sim C_{n} + \textrm{d}\Lambda_{n-1}
\eeq
One may view the $C_n$ as connections on an $(n-1)$-gerbe,
defined over the overlap patches. Their gauge invariance becomes
important on \emph{triple} overlaps. Consistency requires that on such
triple overlaps the $n$--forms of the double overlaps sum to an exact
form, say $\textrm{d}C_{n-1}$. The $C_{n-1}$ have their own gauge
invariance, and can therefore be viewed as connections on $(n-2)$
gerbes, defined over the triple overlaps. The argument repeats until
one reaches the zero forms, which form the transition functions for
$0$-gerbes (that is, line bundles). 

We specify the connection on the $n$-gerbe, by the tower of
$(n+1)$--forms on patches, $n$--forms on overlaps,
$(n-1)$--forms on triple overlaps, etc. We require the gerbe to be
flat, but as with bundles, this doesn't mean the gerbe is
trivial; there may be non-trivial holonomy around closed
$(n+1)$--cycles, and hence we should compute it. For a bundle we would
integrate over patches, and correct for the
transitionfunctions on the overlap patches. The generalization of this
procedure reads as follows\cite{deBoer, Mackaay}
\begin{eqnarray} \label{hol}
\int_M C = \sum_i \int_{M_i} C_i - \sum_{ij} \int_{M_{ij}} C_{ij} +
\sum_{ijk} \int_{M_{ijk}} C_{ijk} - \ldots.
\end{eqnarray}
In words: ``Cut up'' the manifold $M$ in pieces (by a partition of unity)
and integrate the $(n+1)$-forms $C_i$, defined on the patches
$M_i$. On the boundaries $M_{ij}$ connecting the patches $M_i$ and
$M_j$ there is a gauge transformation, specified by the form
$C_{ij}$. Integrate all the $C_{ij}$'s over the $M_{ij}$. On
the triple boundaries $M_{ijk}$ connecting $M_i$, $M_j$ and $M_k$ the
form $C_{ijk}$ specifies the connecting gauge transformation, Integrate
all $C_{ijk}$ over the $M_{ijk}$, etc. In the end, combine all the
partial results in an alternating sum. The first term in eq. \ref{hol}
can be viewed as a ``bulk'' contribution, the second is a correction
due to the overlap patches, the third a correction due triple
overlaps, etc.

Equation \ref{hol} is invariant (for $M$ without boundary) under the
gauge transformations
\begin{eqnarray}
C_i  & \rightarrow & C_i + dL_i \nonumber \\
C_{ij} & \rightarrow & C_{ij} + L_i + L_j + d L_{ij} \\
C_{ijk} & \rightarrow & C_{ijk} + L_{ij} + L_{jk} + L_{ki} + d
L_{ijk}, \textrm{ etc.} \nonumber
\end{eqnarray} 

\section{A classification of orientifolds}

We now return to our problem. The $Z_2$ action on the orientifold
$T^k/Z_2$ gives rise to $2^k$ O$(9-k)$ planes. The identities of these planes
are specified by two discrete fluxes ${\cal B}$ and ${\cal C}$. The question is which configurations of planes give rise to flat
gerbes. For convenience we put coordinates on the $k$-torus with
period $2$. The fixed points of $Z_2$ have coordinates that
are either 0 or 1, and the $O(9-k)$-planes can be viewed as elements of $(Z_2)^k$.

Consider first the ${\cal B}$ flux. This gives rise to a 1-gerbe (or
gerbe for short). The holonomy formula eq. \ref{hol} has 3 terms. Focus on
any of the O$(9-k)$ planes, with coordinates $p_i$, and compute (for a generic flat gerbe) the
holonomy around this plane. The result is \cite{Keurentjes01b}
\beq \label{bform}
{\cal B}(\{p_i \})  \equiv \int_{M_p} B = \sum_{i < j} b_{ij} p_i p_j + 
\sum_i b_i p_i + b. 
\eeq
The three terms in this result come from the separate terms in
eq. \ref{hol}. The $b_{ij}, b_i$ and $b$ are $Z_2$ coefficients,
antisymmetric in their indices, that specify the flat gerbe. All
multiplications and sums on the r.h.s of this equation are
within the field $Z_2$, and hence ${\cal B}$ is also an element of
$Z_2$. Eq. \ref{bform} computes whether a given O$(9-k)$ plane carries
a ${\cal B}$ flux, and therefore partly specifies the identity of the plane.

The formula for the RR $(k-4)$--form depends on the value of $k$. The
result (see \cite{Keurentjes01b}) is similar to that for ${\cal B}$:
A flat $(k-3)$-gerbe over $T^k/Z_2$ is specified by finitely many
coefficients, that specify the ${\cal C}$ flux of each O$(9-k)$ plane.

By letting $b_{ij}$, $b_i$, $b$, and the coefficients for the RR
$(k-3)$--gerbe, take all possible $Z_2$ values, one finds all
configurations of orientifold planes giving rise to flat gerbes. This
is more than we want. There are coordinate transformations in
\beq
(Z_2)^k \times SL(k,Z_2),
\eeq
that map $T^k/Z_2$ to itself, but permute the orientifold planes. We are only interested in different configurations modulo these
coordinate changes. Using the transformations to restrict the
polynomial eq. \ref{bform} and its RR-counterpart to
suitably chosen standard forms, the redundancy is eliminated.

Having specified the gerbes, we indirectly have specified the
identities of all the orientifold planes. These contribute to the
RR-tadpole, which has to be canceled by adding either D-branes, or
anti D-branes. We will demand unbroken
supersymmetry (which is necessary if the configuration is T-dual to a
toroidal compactification of the type I string with flat bundle) and
hence a tadpole that can be canceled by adding D-branes.

The rest of the classification program consists of straightforward but
tedious computations, for which we refer the reader to
\cite{Keurentjes01b}. The detailed results can also be found
there. The following table gives a summary of some results.
\beq \label{tab}
\begin{tabular}{|c||ccccc||c|} \hline 
$k$ & $r=16$ & $r=8$ & $r=4$ & $r=2$ & $r=0$ & total \\
\hline
0 & 1 & 0 & 0 & 0 & 0 & 1 \\
1 & 1 & 0 & 0 & 0 & 1 & 2 \\
2 & 1 & 1 & 0 & 0 & 1 & 3 \\
3 & 1 & 1 & 0 & 0 & 2 & 4 \\
4 & 1 & 2 & 1 & 1 & 2 & 7 \\
5 & 1 & 2 & 2 & 1 & 7 & 13 \\
6 & 1 & 3 & 4 & 6 & 21 & 35 \\
\hline
\end{tabular}
\eeq
The first column specifies the number $k$ in $T^k/Z_2$. The next
columns specify, for a given value of $k$ how many distinct (up to
coordinate transformations) configurations of orientifold planes
exist on $T^k/Z_2$, such that the resulting theory is supersymmetric,
with manifest\footnote{We are ignoring the
gauge bosons resulting from Kaluza-Klein reduction on metric and
forms.} gauge group of rank $r$. The last column gives the
total of different models that were identified. Problems in finding a
standard form for the 3-form have prevented us from going beyond $k=6$ (but see
\cite{Keurentjes01b} for some isolated results).

A point which should be stressed, is that the classification of
\cite{Keurentjes01b} is \emph{not} a classification of components in
the moduli space of string theories with 16 supersymmetries. There are
components in this moduli space that do not allow an orientifold
description, while other components have more than
one orientifold description (see e.g. \cite{deBoer}).

We motivated this work by referring to type I-heterotic
duality. However, not all models we found are T-dual to a type I
compactification on a $k$-torus with a flat bundle. A second class of models
that enters our classification is dual to compactifications of the
``shift orientifold'' of IIB theory. Compactifying the IIB theory on a
circle, one can combine the orientifold projection with a shift over the circle
\cite{Dabholkar, Witten97}, leaving 16 supersymmetries unbroken. This
construction also has a gauge theory interpretation \cite{Keurentjes00a, Keurentjes01}. 

In 4-dimensions, we find $N=4$ supersymmetric gauge theories and
$SL(2,Z)$-duality. Only $SL(2,Z_2)$, which is isomorphic to the
permutation group on 3 elements, is manifest in our set-up. All
theories organize in orbits under the $SL(2,Z_2)$. None of these
orbits has the maximal length (6), most have length 3. There are 2
orientifold configurations that are singlets under $SL(2,Z_2)$, one
corresponding to the standard compactification, and one non-trivial
one with rank 4 gauge group.

A possible criticism on this work could be that it ignores the modern
opinion that K-theory, rather than cohomology is the appropriate
formalism for RR-fluxes, also in an orientifold
set-up \cite{Bergman}. A number of facts
(K-theory gives the same results for isolated planes \cite{Bergman},
many of our models have known heterotic duals \cite{deBoer,
Keurentjes01}, invalidating a single theory would render the $SL(2,Z)$
orbits of 4-d S-duality incomplete) give us confidence that a honest
K-theory classification would not differ much from ours. Our classification may
be a suitable starting point for research in this direction.

{\bf Acknowledgments}: I enjoyed discussing with Jan de Boer, Robbert
Dijkgraaf, Eric Gimon, Amihay Hanany, Bernard Julia and Jussi
Kalkkinen. Niels Keurentjes has helped me with a computer program used
for this research. This work is partly supported by EU contract HPRN-CT-2000-00122. 

%%%%%%%%%%%%%%%%%%%%%%
%%%%%%%%%%%%%%%%%%%%%%

\end{document}